\documentclass[prd,twocolumn,showpacs,amsmath,amssymb]{revtex4}
\usepackage{graphicx}
\usepackage{dcolumn}
\usepackage{bm}

\begin{document}

\title{Power spectrum and anisotropy of super inflation in loop quantum cosmology  }
\author{Xiao-Jun Yue}
\email{yuexiaojun@mail.bnu.edu.cn}
  \affiliation{Department of Physics, Beijing Normal University, Beijing 100875, China}
\author{Jian-Yang Zhu}
\thanks{Author to whom correspondence should be addressed}
\email{zhujy@bnu.edu.cn}
\affiliation{Department of Physics, Beijing Normal University, Beijing 100875, China}
\date{\today}

\begin{abstract}
We investigate the scalar mode of perturbation of super inflation in the version of  loop quantum cosmology in which the gauge invariant holonomy corrections are considered. Given a background solution, we calculate the power spectrum of the perturbation in the classical and LQC conditions. Then we compute the anisotropy originated from the perturbation. It is found that in the presence of the gauge invariant holonomy corrections the power spectrum is exponentially blue and the anisotropy also grows exponentially in the epoch of super inflation.
\end{abstract}

\pacs{98.80.-k,98.80.Cq,98.80.Qc}
\maketitle

\section{\label{s1}Introduction}
The inflation model is the most promising one subject to observational tests in cosmology, which can express the origin of the large scale structure in the Universe \cite{Starobinsky,Guth,Albrecht-Steinhardt,Hawking-Moss,Linde,Linde-Lyth}. In the form of inflation theory, all the structure we see in the Universe is a result of quantum fluctuations during the inflation epoch, including the observed cosmic microwave background (CMB) anisotropy and the large-scale distribution of galaxies and dark matter. However, what fundamental theory  the inflation seems to arise from is still a question. In the simplest versions, inflation is realized by a scalar field, whose kinetic energy is negligible compared to the potential energy. Other theories such as string and M theory can also give rise to inflation. In this paper we focus our attention to the form of loop quantum gravity(LQG) \cite{lqg1,lqg2,lqg3}.

Loop quantum gravity (LQG) is a background independent and nonperturbative canonical quantization of general relativity. The variables used here are the  holonomies of the connection and fluxes of the triad. Its cosmological version, the loop quantum cosmology (LQC) \cite{B1-B4}, which is the application of LQG to symmetric states (see Refs.\cite{Rovelli-LivingRevRel,Thiemann-Lect,Bojowald-grqc}), has achieved many successes. In particular, LQC can lead to a nonsingular evolution of the Universe \cite{Bojowald-PRL-01}, with the behavior being traced to the discreteness of the spacetime. Although the scheme of the  evolution is fascinating, the rigorous approach is very difficult to afford and it is not easy to connect it with the existing theories based on classical dynamics. In dealing with this problem, another approach  based on the effective or semiclassical equations comes out. Given modifications of the discreteness of spacetime to the classical dynamics, a number of important quantum effects can arise. There are two types of corrections that are expected from the Hamiltonian of  LQG. One is the¡°inverse volume correction¡± and the other is the ¡°holonomy correction¡±. With the inverse volume correction, the classical dynamics is modified to include high energy corrections that originate from the specrtra of quantum operators related to the inverse scale factor \cite{Bojowald-PRD-01,Bojowald-CQG-02,Vandersloot-PRD-05}. The holonomy correction is due to the use of holonomies as a basic variable in the quantization scheme \cite{APS-PRL06-PRD06,APSV-PRD07,Vandersloot-PRD07,CVZ-PRD07}.

One of the most interesting quantum effects due to LQC is the presence of a super inflation epoch which occurs during the early phase of the Universe after the big bounce, independently of the potential \cite{Bojowald-PRL02,BV-PRD03}. In classical dynamics, the standard inflation is driven by a self-interacting potential and super inflation cannot happen unless bringing in the exotic matter, which violates the null energy condition (NEC). However, in LQC it is shown that the super inflation period will happen independently of the form of the potential \cite{DH-PRL05}. LQC predicts an era of super inflation irrespective of whether it is followed by a standard slow-roll inflation or not.

Because of the robustness of the super inflation in LQC, it makes great sense to study the perturbation and anisotropy given by such a phase. The application of the scalar mode perturbation with inverse volume correction can be found in \cite{lqcs}, the vector mode in \cite {lqcv}, and the tensor mode in \cite{lqct}. However, when the perturbation of inhomogeneities  around the Friedmann-Robertson-Walker spacetime is investigated the anomaly problem appears. When attempting to include the corrections of LQC, the constraints with the presence of quantum corrections are not anomaly-free, which means the derived background equations can no longer be compatible with all the terms in the perturbation equation. This problem is solved in the ¡°inverse volume correction¡± scenario by adding counterterms to the constraints to eliminate the anomalous terms and making the constraint algebra to be closed \cite{lqcg1}. The gauge invariant cosmological perturbation equations are also derived in \cite{BHKS-PRD09}.

Scalar perturbation with holonomy corrections have also been studied in \cite{Wu-Ling}, and the power spectrum of the perturbed scalar field is also calculated in\cite{CMNS-PRD08}. However, most of the previous investigations are based on calculations which are not anomaly free and the power spectrum calculated is also under the approximation that the background spacetime is unperturbed and only the scalar field is perturbed. Recently, the gauge-invariant scalar mode perturbation with holonomy corrections has been found in \cite{afh}, where it is also obtained by introducing counterterms to eliminate the anomaly problem as in the inverse volume correction case, and the equations of perturbation are also derived in that paper.

With the quantum effects described by LQC, it is natural to ask whether the signatures of the super inflation period can be obtained. To answer this question, we have to consider the power spectrum and anisotropy produced by this epoch. In this paper we focus our attention to the scalar mode perturbation with the gauge-invariant form of holonomy corrections. We calculate the power spectrum and anisotropy risen by a scalar field in classical dynamics and in LQC with holonomy corrections, respectively, and find the quantum effects in this period.

This paper is organized as follows. The gauge-invariant scalar mode perturbation with holonomy corrections is presented in Sec.\ref{s2}. In Sec.\ref{s3} and Sec. \ref{s4} we calculate, respectively, the power spectrum of the perturbation during the super inflation epoch and the anisotropy during this period. At last, in Sec. \ref{s5}, we give some discussion and conclusion.

\section{\label{s2}Scalar perturbation of LQC}

In this section, we review briefly the gauge-invariant formalism of the cosmological perturbation theory with holonomy correction in LQC. The
detailed derivation can be found in \cite{afh}. For the scalar mode perturbations, along with the background FRW metric, the line element takes the form
\begin{eqnarray}
ds^2 &=&a^2(\tau )\left\{ -\left( 1+2\phi \right) d\tau ^2+2\partial
_aBd\tau dx^a\right.   \nonumber \\
&&\left. +\left[ \left( 1-2\psi \right) \delta _{ab}+2\partial _a\partial
_bE\right] dx^adx^b\right\} ,  \label{FRWmetric}
\end{eqnarray}
where the scale factor $a$ is a function of the conformal time $\tau $, the spatial indices $a$ and $b$ run from 1 to 3, and $\phi $, $\psi $, $E$, and $B$ are scalar perturbation functions.

In the Hamiltonian framework, the background variables are $\bar{k}$, $\bar{p}$, $\bar{\varphi}$, and $\bar{\pi}$, while the perturbed variables are $\delta K_a^i$, $\delta E_i^a$, $\delta \varphi $, and $\delta \pi $. When the holonomy corrections are taking into account in the scalar mode perturbation, we have to add some counterterms in the Hamiltonian to eliminate the anomalies. The explicit form of holonomy-modified gravitational Hamiltonian constraint can be written as
\begin{equation}
H_G^Q[N]=\frac 1{2\kappa }\int_\Sigma d^3x\left[ \bar{N}\left( {\cal H}%
_G^{(0)}+{\cal H}_G^{(2)}\right) +\delta N{\cal H}_G^{(1)}\right] ,
\label{Hamiltonian(G)}
\end{equation}
where
\[
{\cal H}_G^{(0)}=-6\sqrt{\bar{p}}\left( \mathbb{K}[1]\right) ^2,
\]
\begin{eqnarray*}
{\cal H}_G^{(1)} &=&-4\sqrt{\bar{p}}\left( \mathbb{K}[s_1]+\alpha _1\right)
\delta _j^c\delta K_c^j-\frac 1{\sqrt{\bar{p}}} \\
&&\times \left( \mathbb{K}[1]^2+\alpha _2\right) \delta _c^j\delta E_j^c+%
\frac 2{\sqrt{\bar{p}}}\left( 1+\alpha _3\right) \partial _c\partial
^j\delta E_j^c,
\end{eqnarray*}
\begin{eqnarray*}
{\cal H}_G^{(2)} &=&\sqrt{\bar{p}}\left( 1+\alpha _4\right) \delta
K_c^j\delta K_d^k\delta _k^c\delta _j^d \\
&&-\sqrt{\bar{p}}\left( 1+\alpha _5\right) \left( \delta K_c^j\delta
_j^c\right) ^2 \\
&&-\frac 2{\sqrt{\bar{p}}}\left( \mathbb{K}[s_2]+\alpha _6\right) \delta
E_j^c\delta K_c^j \\
&&-\frac 1{2\bar{p}^{3/2}}\left( \mathbb{K}[1]^2+\alpha _7\right) \delta
E_j^c\delta E_k^d\delta _c^k\delta _d^j \\
&&+\frac 1{4\bar{p}^{3/2}}\left( \mathbb{K}[1]^2+\alpha _8\right) \left(
\delta E_j^c\delta _c^j\right) ^2 \\
&&-\frac 1{2\bar{p}^{3/2}}\delta ^{jk}\left( \partial _c\delta E_j^c\right)
\left( \partial _d\delta E_d^k\right) ,
\end{eqnarray*}
and, in the above expression,
\[
\mathbb{K}[n]=\frac{sin(n\bar{\mu}\gamma \bar{k})}{n\bar{\mu}\gamma },\ %
\mathbb{K}[0]=\bar{k}.
\]
The $\alpha_i$ are counterterms, which are introduced to remove anomalies and vanish in the classical limit $(\bar{\mu}\rightarrow 0)$. In this paper
we introduce the matter to be a scalar field, and the scalar matter Hamiltonian can be expressed as
\begin{equation}
H_M^Q\left[ N\right] =H_M[\bar{N}]+H_M[\delta N]  \label{Hamiltonian(M)},
\end{equation}
where
\begin{eqnarray*}
H_M[\bar{N}] &=&\int_\Sigma d^3x\bar{N}\left[ \left( {\cal H}_\pi ^{(0)}+%
{\cal H}_\varphi ^{(0)}\right) \right.  \\
&&\left. +\left( {\cal H}_\pi ^{(2)}+{\cal H}_\nabla ^{(2)}+{\cal H}_\varphi
^{(2)}\right) \right],
\end{eqnarray*}
and
\[
H_M[\delta N]=\int_\Sigma d^3x\delta N\left[ {\cal H}_\pi ^{(1)}+{\cal H}%
_\varphi ^{(1)}\right].
\]
The factors in the above equations are, respectively
\[
{\cal H}_\pi ^{(0)}=\frac{\bar{\pi}^2}{2\bar{p}^{3/2}},{\cal H}_\varphi
^{(0)}=\bar{p}^{3/2}V(\bar{\varphi}),
\]
\begin{eqnarray*}
{\cal H}_\pi ^{(2)} &=&\frac 12\frac{\delta \pi ^2}{\bar{p}^{3/2}}-\frac{%
\bar{\pi \delta \pi }}{\bar{p}^{3/2}}\frac{\delta _c^j\delta E_j^c}{2\bar{p}}
\\
&&+\frac 12\frac{\bar{\pi}^2}{\bar{p}^{3/2}}\left[ \frac{(\delta _c^j\delta
E_j^c)^2}{8\bar{p}^2}+\frac{\delta _c^k\delta _d^j\delta E_j^c\delta E_k^d}{4%
\bar{p}^2}\right] ,
\end{eqnarray*}
\[
{\cal H}_\nabla ^{(2)}=\frac 12\sqrt{\bar{p}}(1+\alpha _{10})\delta
^{ab}\partial _a\delta \varphi \partial _b\delta \varphi ,
\]
\begin{eqnarray*}
{\cal H}_\varphi ^{(2)} &=&\frac 12\bar{p}^{3/2}V_{,\varphi \varphi }(\bar{%
\varphi})\delta \varphi ^2+\bar{p}^{3/2}V,_\varphi (\bar{\varphi})\delta
\varphi \frac{\delta _c^j\delta E_j^c}{2\bar{p}} \\
&&+\bar{p}^{3/2}V(\bar{\varphi})\left[ \frac{(\delta _c^j\delta E_j^c)^2}{8%
\bar{p}^2}-\frac{\delta _c^k\delta _d^j\delta E_j^c\delta E_k^d}{4\bar{p}^2}%
\right] ,
\end{eqnarray*}
\[
{\cal H}_\pi ^{(1)}=\frac{\bar{\pi}\delta \pi }{\bar{p}^{3/2}}-\frac{\bar{\pi%
}^2}{2\bar{p}^{3/2}}\frac{\delta _c^j\delta E_j^c}{2\bar{p}},
\]
and
\[
{\cal H}_\varphi ^{(1)}=\bar{p}^{3/2}\left[ V_{,\varphi }(\bar{\varphi}%
)\delta \varphi +V(\bar{\varphi})\frac{\delta _c^j\delta E_j^c}{2\bar{p}}%
\right] .
\]
The counter-terms $\alpha _i$ are given by
\[
\left\{
\begin{array}{c}
\alpha _1=\mathbb{K}[2]-\mathbb{K}[s_1], \\
\alpha _2=2\mathbb{K}[1]^2-2\bar{k}\mathbb{K}[2], \\
\alpha _6=2\mathbb{K}[2]-\mathbb{K}[s_2]-\bar{k}\Omega , \\
\alpha _7=-4\mathbb{K}[1]^2+6\bar{k}\mathbb{K}[2]-2\bar{k}^2\Omega , \\
\alpha _8=-4\mathbb{K}[1]^2+6\bar{k}\mathbb{K}[2]-2\bar{k}^2\Omega \\
\alpha _4=\alpha _5=\alpha _{10}=\Omega -1, \\
\alpha _3=\alpha _9=0,
\end{array}
\right.
\]
and $\Omega $ is defined as
\begin{equation}
\Omega :=\cos \left( 2\bar{\mu}\gamma \bar{k}\right) =1-\frac{2\rho }{\rho _c%
}  \label{Omega}
\end{equation}
where $\rho _c$ is the critical energy density.

\subsection{\label{s2a}Background equations}

Using the anomaly-free conditions \cite{afh}, we can get the equations of motion for the canonical variables through the Hamilton
equation
\begin{equation}
f^{\prime }=\left\{ f,H\left[ N,N^a\right] \right\},  \nonumber
\end{equation}
where the Hamiltonian $H[N,N^a]$ is the sum of all constraints
\begin{equation}
H[N,N^a]=H_G^Q[N]+H_M[N]+D_G\left[ N^a\right] +D_M\left[ N^a\right].
\label{Hamiltonian}
\end{equation}
The equations of the background variables are the following:
\begin{eqnarray}
\bar{k}^{\prime }&=&-\frac{\bar{N}}{2\sqrt{\bar{p}}}\mathbb{K}[1]^2-\bar{N}%
\sqrt{\bar{p}}\frac \partial {\partial \bar{p}}\mathbb{K}[1]^2  \nonumber\\
&&+\frac \kappa 2\sqrt{\bar{p}}\bar{N}\left[ -\frac{\sqrt{\bar{\pi}}^2}{2\bar{p}^3}+V(\varphi
)\right] ,  \label{k'}
\end{eqnarray}
\begin{equation}
\bar{p}^{\prime }=2\bar{N}\sqrt{\bar{p}}\mathbb{K}[2],  \label{p'}
\end{equation}
\begin{equation}
\bar{\varphi}^{\prime }=\bar{N}\frac{\bar{\pi}}{\bar{p}^{3/2}},  \label{phi'}
\end{equation}
\begin{equation}
\bar{\pi}^{\prime }=-\bar{N}\bar{p}^{3/2}V,_\varphi (\bar{\varphi}),
\label{pi'}
\end{equation}
where the ¡±$^{\prime }$¡± means the differentiation with the conformal time $\tau $.

Now we set $\bar{N}=\sqrt{\bar{p}}$, and then we can combine Eqs.(\ref{phi'}) and (\ref{pi'}) to the Klein-Gordon equation
\begin{equation}
\bar{\varphi}^{\prime \prime }+2\mathbb{K}[2]\bar{\varphi}^{\prime }+\bar{p}%
V,_\varphi (\bar{\varphi})=0.  \label{Klein-Gordon}
\end{equation}
In addition, Eq.(\ref{p'}) can lead to the modified Friedmann equation
\begin{equation}
{\cal H}^2=\bar{p}\frac \kappa 3\rho \left( 1-\frac \rho {\rho _c}\right) ,
\label{Hub^2}
\end{equation}
where ${\cal H}$ is the conformal Hubble rate and has the expression
\begin{equation}
{\cal H}=\frac{\bar{p}^{\prime }}{2\bar{p}}=\mathbb{K}[2].  \label{Hub}
\end{equation}
Another useful expression is
\begin{equation}
3\mathbb{K}[1]^2=\kappa \left( \frac{\bar{\pi}}{2\bar{p}^2}+\bar{p}V(\bar{%
\varphi})\right) .
\end{equation}
In this paper, we consider the situation in LQC that the matter is introduced as
a normal scalar field which satisfies the energy condition, in which the energy
density and pressure are given by, respectively,
\begin{equation}
\rho =\frac{\bar{\pi}^2}{2\bar{p}^3}+V(\varphi ),\ P=\frac{\bar{\pi}^2}{2%
\bar{p}^3}-V(\varphi ).  \nonumber
\end{equation}

\subsection{\label{s2b}Perturbation equations}

The equations for the perturbed parts of the canonical variables are, respectively
\begin{eqnarray}
\delta E_i^{a^{\prime }} &=&-\bar{N}\left[ \sqrt{\bar{p}}\Omega \delta
K_c^i\delta _i^c\delta _j^a-\sqrt{\bar{p}}\Omega \left( \delta K_c^j\delta
_j^c\right) \delta _i^a\right.   \nonumber \\
&&\left. -\frac 1{\sqrt{\bar{p}}}\left( 2\mathbb{K}[2]-\bar{k}\Omega \right)
\delta E_i^a\right] +\delta N\left( 2\mathbb{K}\sqrt{\bar{p}}\delta
_i^a\right)   \nonumber \\
&&-\bar{p}\left[ \partial _i\delta N^a-\left( \partial _c\delta N^c\right)
\delta _i^a\right] ,  \label{delta E'}
\end{eqnarray}
\begin{eqnarray}
\delta K_a^{i^{\prime }} &=&\bar{N}\left[ -\frac 1{\sqrt{\bar{p}}}\left( 2%
\mathbb{K}[2]-\bar{k}\Omega \right) \delta K_a^i+\frac{\delta ^{ik}}{2\bar{p}%
^{3/2}}\partial _a\partial _d\delta E_k^d\right.   \nonumber \\
&&-\frac 1{2\bar{p}^{3/2}}\left( -3\mathbb{K}[1]^2+6\bar{k}\mathbb{K}[2]-2%
\bar{k}^2\Omega \right) \delta E_j^c\delta _a^j\delta _c^i  \nonumber \\
&&\left. +\frac 1{4\bar{p}^{3/2}}\left( -3\mathbb{K}[1]^2+6\bar{k}%
\mathbb{K}[2]-2\bar{k}^2\Omega \right) \left( \delta E_j^c\delta _c^j\right)
\delta _a^i\right]   \nonumber \\
&&+\frac 12\left[ \frac 2{\sqrt{\bar{p}}}\left( \partial _a\partial ^i\delta
N\right) -\frac 1{\sqrt{\bar{p}}}\left( 3\mathbb{K}[1]^2-2\bar{k}%
\mathbb{K}[2]\right) \delta _a^i\delta N\right]   \nonumber \\
&&++\delta _c^i\left( \partial _a\delta N^c\right) \kappa \delta N\frac{%
\sqrt{\bar{p}}}2\left( -\frac{\bar{\pi}^2}{2\bar{p}^{3/2}}+V\left( \bar{%
\varphi}\right) \right) \delta _a^i  \nonumber \\
&&+\kappa \bar{N}\left[ -\frac{\bar{\pi}\delta \pi }{2\bar{p}^{5/2}}\delta
_a^i+\frac{\sqrt{\bar{p}}}2\delta \varphi \frac{\partial V(\bar{\varphi})}{%
\partial \bar{\varphi}}\delta _a^i\right.   \nonumber \\
&&+\left( \frac{\bar{\pi}^2}{2\bar{p}^{3/2}+\bar{p}^{3/2}}V(\varphi )\right)
\frac{\delta _c^i\delta E_j^c}{4\bar{p}^2}\delta _a^i  \nonumber \\
&&\left. +\left( \frac{\bar{\pi}^2}{2\bar{p}^{3/2}}-\bar{p}^{3/2}V(\varphi
)\right) \frac{\delta _c^i\delta _a^j\delta E_j^c}{2\bar{p}^2}\right] ,
\label{delta K'}
\end{eqnarray}
\begin{equation}
\delta \varphi ^{\prime }=\delta N\left( \frac{\bar{\pi}}{\bar{p}^{3/2}}%
\right) +\bar{N}\left( \frac{\delta \pi }{\bar{p}^{3/2}}-\frac{\bar{\pi}}{%
\bar{p}^{3/2}}\frac{\delta _c^j\delta E_j^c}{2\bar{p}}\right) ,
\label{delta phi'}
\end{equation}
and
\begin{eqnarray}
\delta \pi ^{\prime } &=&-\delta N\left[ \bar{p}^{3/2}V,_\varphi \left( \bar{%
\varphi}\right) +\bar{\pi}\left( \partial _a\delta N^a\right) \right]
\nonumber \\
&&-\bar{N}\left[ -\sqrt{\bar{p}}\Omega \delta ^{ab}\partial _a\partial
_b\delta \varphi \right.   \nonumber \\
&&\left. +\bar{p}^{3/2}V,_\varphi \left( \bar{\varphi \varphi }\right)
\left( \bar{\varphi}\right) \frac{\delta _c^j\delta E_j^c}{2\bar{p}}\right] ,
\label{delta pi'}
\end{eqnarray}
where
\[
\left\{
\begin{array}{c}
\delta E_i^a=-2\bar{p}\psi \delta _i^a+\bar{p}(\delta _i^a\triangle
-\partial ^a\partial _i)E, \\
\delta N=\sqrt{\bar{p}}\phi ,\ \delta N^a=\partial ^aB.
\end{array}
\right.
\]

Applying the above expressions to the left of Eq. (\ref{delta E'}), one can get the following expression
\begin{equation}
\delta K_a^i=-\frac 1\Omega (\psi ^{\prime }+k\Omega \psi +{\cal H}\phi
)\delta _a^i+\partial _a\partial ^i(\bar{k}E-\frac 1\Omega (B-E^{\prime })).
\label{delta K}
\end{equation}
Furthermore, applying Eq.(\ref{delta K'}) to the left-hand side of Eq.(\ref{delta phi'}), one
can also get two other equations, which correspond to the diagonal and
off-diagonal parts:
\begin{eqnarray}
&&\psi ^{\prime \prime }+2\left( {\cal H}-\frac{\Omega ^{\prime }}\Omega
\right) \psi ^{\prime }+{\cal H}\phi ^{\prime }+\left( 2{\cal H}^2+{\cal H}%
^{\prime }-{\cal H}\frac{\Omega ^{\prime }}\Omega \right) \phi   \nonumber \\
&=&\frac \kappa 2\Omega \left( \bar{\varphi}^{\prime }\delta \varphi
^{\prime }-\bar{p}V,_\varphi \delta \varphi \right) ,
\end{eqnarray}
\begin{equation}
\psi -\phi +(\frac{\Omega ^{\prime }}\Omega -\frac{2{\cal H}}\Omega
)(B-E^{\prime })-\frac 1\Omega {(B-E^{\prime })^{\prime }}=0.
\end{equation}

The following diffeomorphism constraint equation
\begin{eqnarray}
0 &=&\kappa \frac{\delta H\left[ N,N^a\right] }{\delta \left( \delta
N^c\right) }=\bar{p}\partial _c\left( \delta _k^d\delta K_d^k\right) -\bar{p}%
\left( \partial _k\delta K_c^k\right)  \\
&&-\bar{k}\delta _c^k\left( \partial _d\delta E_k^d\right) +\kappa \bar{\pi}%
\left( \partial _c\delta \varphi \right)
\end{eqnarray}
can be expressed equivalently as
\begin{equation}
\partial _c\left[ \psi ^{\prime }+{\cal H}\phi \right] =4\pi G\bar{\varphi}%
^{\prime }\partial _c\delta \varphi .
\label{PerturbationEq-1}
\end{equation}
And, in addition, with the expressions of $\delta K_a^i$ and $\delta E_i^a$, the perturbed part of the Hamiltonian constraint,
\begin{eqnarray}
0 &=&\frac{\delta H[N,N^a]}{\delta (\delta N)}=\frac 1{2\kappa }\left[ -4%
\sqrt{\bar{p}}\mathbb{K}[2]\delta _j^c\delta K_c^j\right.   \nonumber \\
&&\left. -\frac 1{\sqrt{\bar{p}}}\left( 3\mathbb{K}[1]^2-2\bar{k}{\cal H}%
\right) \delta _c^j\delta E_j^c+\frac 2{\sqrt{\bar{p}}}\partial _c\partial
^i\delta E_j^c\right]   \nonumber \\
&&+\frac{\bar{\pi}\delta \pi }{\bar{p}^{3/2}}-\frac{\bar{\pi}^2}{2\bar{p}%
^{3/2}}\frac{\delta _c^j\delta E_j^c}{2\bar{p}}  \nonumber \\
&&+\bar{p}^{3/2}\left[ V,_\varphi \left( \bar{\varphi}\delta \varphi +V(\bar{%
\varphi})\frac{\delta _c^j\delta E_j^c}{2\bar{p}}\right) \right] ,
\end{eqnarray}
can be written equivalently as
\begin{eqnarray}
&&\Omega \nabla ^2\psi -3{\cal H}\psi ^{\prime }-\left( 2{\cal H}^2+{\cal H}%
^{\prime }\right) \phi -{\cal H}\nabla ^2\left( B-E^{\prime }\right)
\nonumber \\
&=&4\pi G\Omega \left( \bar{\varphi}^{\prime }\delta \varphi ^{\prime }+\bar{%
p}V,_\varphi \delta \varphi \right).   \label{PerturbationEq-2}
\end{eqnarray}
Finally, we give out the gauge transformation forms with scalar
perturbations under a small coordinate transformation
\[
x^\mu \rightarrow x^\mu +\xi ^\mu ;\xi ^\mu =\left( \xi ^0,\partial ^a\xi
\right) .
\]
Taking into account the holonomy corrections, the transformations of the metric perturbations can be expressed as follows
\[
\left\{
\begin{array}{c}
\delta _{\left[ \xi ^0,\xi \right] }\psi =-{\cal H}\psi ^0, \quad \quad \\
\delta _{\left[ \xi ^0,\xi \right] }\phi =\xi ^{0^{\prime }}+{\cal H}\xi ^0,
\\
\delta _{\left[ \xi ^0,\xi \right] }E=\xi ,  \quad \quad \quad \quad\\
\delta _{\left[ \xi ^0,\xi \right] }B=\xi ^{\prime }, \quad \quad \quad \label{26}
\end{array}
\right.
\]
and the following gauge transformations of the time derivative of a variable $X$ can also be expressed as
\begin{equation}
\delta _{\left[ \xi ^0,\xi \right] }X^{\prime }-\left( \delta _{\left[ \xi
^0,\xi \right] }X\right) ^{\prime }=\Omega \cdot \delta _{\left[ 0,\xi
^0\right] }X.
\end{equation}

It is possible to define the Mukhanov variable as follow
\begin{equation}
v:=\sqrt{\bar{p}}\left( \delta \varphi +\frac{\bar{\varphi}^{\prime }}{%
\mathbb{K}[2]}\psi \right) =a(\eta )\left( \delta \varphi +\frac{\bar{\varphi%
}^{\prime }}{{\cal H}}\psi \right),   \label{Mukhanov v}
\end{equation}
and we can derive the corresponding dynamical equation
\begin{equation}
v^{\prime \prime }-\Omega \nabla ^2v-\frac{z^{\prime \prime }}zv=0,
\label{Mukhanov v''}
\end{equation}
where
\begin{equation}
z=\sqrt{\bar{p}}\frac{\bar{\varphi}^{\prime }}{\mathbb{K}[2]}=a(\eta )\frac{%
\bar{\varphi}^{\prime }}{{\cal H}}.  \label{z}
\end{equation}
It is also possible to define the perturbation of curvature ${\cal R}$ such
that
\begin{equation}
{\cal R}=\frac vz.  \label{R}
\end{equation}

\section{\label{s3}Power spectrum}
In this section we calculate the power spectrum of the perturbed field. As a comparison we first calculate the spectrum in the classical condition, then
we investigate the power spectrum in the LQC form with the holonomy corrections.

\subsection{\label{s11}Power spectrum in the classical condition}

In the classical condition, the time derivative of the Hubble parameter is
\begin{equation}
\dot{H}=-4\pi G(\rho +P)
\end{equation}
where $\rho $ is the total energy density and $P$ is the pressure. We can see from the above equation that the super inflation cannot happen in the classical condition unless the null energy condition (NEC) is violated:
\begin{equation}
\rho +P<0.
\end{equation}
As the simplest example of super inflation we consider the case with an exponential potential:
\begin{equation}
\rho =\frac 12\sigma _K\dot{\varphi}^2+\sigma _VV_0e^{-\lambda \varphi
/M_{pl}}  \label{exp-poten}
\end{equation}
where $\sigma _K$, $\sigma _V=\pm 1$. The case with $\sigma _K=\sigma _V=-1$ leads to solutions in Euclidean time which are not considered here. The case with negative potential ($\sigma _K=-\sigma _V=1$) leads to the simplest single field realization of the Ekpyrotic scenario \cite{KOST-PRD01}. The case $\sigma_K=-\sigma _V=-1$ leads to a stage of the super inflation solution which we will consider here:
\begin{equation}
\left\{
\begin{array}{c}
a(t)\sim (-t)^m,t<0,m<0, \\
\varphi (t)=\frac 2\lambda M_{pl}\log \left( -M_{pl}t\right), \\
V_0=M_{pl}^4m\left( 3m-1\right), \quad \quad \quad
\end{array}
\right.  \label{super-inflation}
\end{equation}
where $m=-2/\lambda ^2$. In the standard super inflation, the Hubble parameter $H$ increases rapidly, at a rate which is much faster than the scale factor $a$. Such a solution
is characterized by a constant state parameter $w=-1+2/(3m)<1$ and $H=m/t>0$, where super inflation happens.

We can also write the solution in the conformal time $\tau $:
\begin{equation}
\left\{
\begin{array}{c}
a(\tau )\sim (-\tau )^p, \quad \quad \quad \quad \quad \quad \quad \quad \\
\varphi (\tau )=\frac{2\left( 1+p\right) }\lambda M_{pl}\log \left(
-M_{pl}\tau \right),  \\
V_0=M_{pl}^4\frac{p\left( 2p-1\right) }{\left( 1+p\right) ^2},\quad \quad \quad \quad \quad \quad
\end{array}
\right.   \label{super-inflation-tau}
\end{equation}
where $p=\frac m{1-m}$ and $p<0$. From the mass conservation equation
\begin{equation}
\dot{\rho}+3H(\rho +P)=0,  \label{mass-conservation}
\end{equation}
and the form of energy density Eq.(\ref{exp-poten}), we can obtain the equation of the homogeneous background scalar field
\begin{equation}
\bar{\varphi}^{\prime \prime }+2{\cal H}\bar{\varphi}^{\prime }-V,_\varphi (%
\bar{\varphi})=0.  \label{K-G-class}
\end{equation}
From the perturbed Einstein equation $\delta G_{\mu \nu }=\delta T_{\mu \nu }$ we can get the equations of the scalar perturbations:
\begin{eqnarray}
&&\frac 2{a^2}\left\{ 3{\cal H}(\psi^{\prime}+{\cal H}\phi )\right.   \nonumber \\
&&\left. -\nabla ^2\left[ \psi +{\cal H}\left( E^{\prime }-B\right) \right]
+\left( {\cal H}^{\prime }-{\cal H}^2\right) \phi \right\}   \nonumber \\
&=&-\delta \rho   \label{perturbation-classEq1}
\end{eqnarray}
\begin{equation}
\frac 2{a^2}\left[ -\left( \psi ^{\prime }+{\cal H}\phi \right) \right]
=\left( \rho +P\right) \delta u , \label{perturbation-classEq2}
\end{equation}
\begin{equation}
\frac 2{a^2}\left[ \psi -\phi +\left( E-B^{\prime }\right) ^{\prime }+2{\cal %
H}\left( E^{\prime }-B\right) \right] =0,  \label{perturbation-classEq3}
\end{equation}
\begin{equation}
\frac 2{a^2}\left[ \left( \psi ^{\prime }+{\cal H}\phi \right) ^{\prime }+2%
{\cal H}\left( \psi ^{\prime }+{\cal H}\phi \right) \right] =\delta P,
\label{perturbation-classEq4}
\end{equation}
where $\delta u=-\frac{\delta \varphi }{\varphi ^{\prime }}$ and the expression of $\delta \rho $ and $\delta P$ can be derived from Eq.(\ref{exp-poten}).

We know that the intrinsic spatial curvature of hypersurface on constant conformal time $\tau $ is
\begin{equation}
R^{(3)}=\frac 4{a^2}\nabla ^2\psi,   \nonumber
\end{equation}
and the variable ${\cal R}$ is the curvature perturbation on thecomoving hypersurface, where $\delta \varphi =0,$ i.e.,
\[
{\cal R}=\psi |_{\delta \varphi =0}=\psi +{\cal H}\frac{\delta \varphi }{\bar{\varphi}^{\prime }}.
\]
By defining
\begin{equation}
v=z{\cal R}  \label{v}
\end{equation}
[see Eq. (\ref{R})], where
\[
z=a(t)\frac{\bar{\varphi}}H=a(\tau )\frac{\varphi ^{\prime }}{{\cal H}},
\]
we can derive the Mukhanov equation from Eqs.(\ref{perturbation-classEq1})-(\ref{perturbation-classEq4}):
\begin{equation}
v^{\prime \prime }-\nabla ^2v-\frac{z^{\prime \prime }}zv=0.
\label{Mukhanov eq}
\end{equation}
Applying the solution equation(\ref{super-inflation-tau}) to the above equation and going to the Fourier space, we obtain
\begin{equation}
v_k^{\prime \prime }+(k^2+m_{eff}^2)v_k=0 , \nonumber
\end{equation}
where
\begin{equation}
m_{eff}^2=-\frac 1{\tau ^2}\left( \nu ^2-\frac 14\right),   \nonumber
\end{equation}
and
\begin{equation}
\nu ^2=p\left( p-1\right) +\frac 14 . \nonumber
\end{equation}
The general solution is
\begin{equation}
v_k=\sqrt{x}\left[ AH_\nu ^{(1)}(x)+BH_\nu ^{(2)}(x)\right]   \nonumber
\end{equation}
where $A$ and $B$ are constants and $H_\nu ^{(1)}$,$H_\nu ^{(2)}$are Hankel's functions of the first and second kind, and $x=-k\tau $. If we impose that in
the ultraviolet regime $k\gg aH$($-k\tau \gg 1$) the solution matches the plane wave solution $e^{-ik\tau }/\sqrt{2k}$ that we expect in flat space
time. On the other hand, the asymptotic behavior of Hankel functions is
\begin{equation}
H_\nu ^{(1)}(x)\sim \sqrt{\frac 2{\pi x}}e^{i\left( x-\frac \pi 2\nu -\frac %
\pi 4\right) },\ x\gg 1  \nonumber
\end{equation}
\[
\quad H_\nu ^{(2)}(x)\sim \sqrt{\frac 2{\pi x}}e^{-i\left( x-\frac \pi 2\nu
-\frac \pi 4\right) },\ x\gg 1
\]
We have to set $B=0$ and $A=\frac{\sqrt{\pi }}2e^{i\left( \nu +\frac 12%
\right) \frac \pi 2}$. Therefore, we have
\begin{equation}
v_k=\frac{\sqrt{\pi }}2e^{i\left( \nu +\frac 12\right) \frac \pi 2}\sqrt{%
-\tau }H_\nu ^{(1)}\left( -k\tau \right).   \label{v_k}
\end{equation}
When the mode gets to the horizon, the Hankel function is $H_\nu^{(1)} \sim \sqrt{2/\pi }e^{-i\frac \pi 2}2^{\nu -\frac 3%
2}x^{-\nu }\Gamma (\nu )/\Gamma (3/2)$, so the function (\ref{v_k}) becomes
\begin{equation}
v_k=e^{i\left( \nu -\frac 12\right) \frac \pi 2}2^{\left( \nu -\frac 32%
\right) }\frac{\Gamma (\nu )}{\Gamma (3/2)}\frac 1{\sqrt{2k}}\left( -k\tau
\right) ^{\frac 12-\nu }.
\end{equation}
Neglecting the constant phase factor, we can get from the definition equation (\ref{v})
\begin{equation}
{\cal R}_k=\frac{\lambda p}{2\sqrt{2}M_{pl}(1+p)}2^{(\nu -\frac 32)}\frac{%
\Gamma (\nu )}{\Gamma (3/2)}k^{-\nu }(-\tau )^{\frac 12-\nu -p}.  \label{R_k}
\end{equation}
The resulting power spectrum is as follows:
\begin{equation}
{\cal P_{{\cal R}}}=\frac{k^3}{2\pi ^2}|{\cal R}|^2\sim k^{3-2\nu }(-\tau
)^{1-2\nu -2p}  \label{P_R}
\end{equation}
If the power spectrum is scale invariant, we have to require $\nu =\frac 32$ which leads to $p=-1$. However, the relation that $m=\frac 1{1+p}$ require $p
$ not to be $-1$. The spectra is a little red tilt and we can choose $p=-1-\delta $ where $\delta $ is a small positive quantity. With this the
spectral index $n_{{\cal R}}$ is
\begin{equation}
\Delta n_{{\cal R}}=3-2\nu =-2\delta.   \nonumber
\end{equation}
Although in the classical condition with the matter violating the null energy condition, the super inflation epoch can lead to a nearly
scale-invariant power spectrum with a little red tilt, the solution is not the standard super inflation during which the Hubble factor rapidly
increases that is much faster than the growth rate of the scale factor. In this condition, $p\rightarrow 0$ and the parameter $\nu $ can be expanded in p: $\nu \approx
\frac 12-p$. The power spectrum in this condition is
\begin{equation}
{\cal P_{{\cal R}}}\sim k^{\Delta n_{{\cal R}}}  \label{P_R-Class}
\end{equation}
and the index
\begin{equation}
\Delta n_{{\cal R}}=2\left( 1+p\right).
\end{equation}

\subsection{\label{s12}Power spectrum in LQC with holonomy corrections}

Given the gauge-invariant form of scalar perturbation with holonomy corrections in the previous section, we are now ready to calculate the spectrum
during the super inflation epoch.

The energy and pressure of the scalar field are given by
\begin{equation}
\rho =\frac{\dot{\varphi}^2}2+V(\varphi ),\ P=\frac{\dot{\varphi}^2}2%
-V(\varphi ) , \nonumber
\end{equation}
which is normal matter in contrast with the classical case. Given this, the time derivative of the Hubble rate is
\begin{equation}
\dot{H}=-\frac{\dot{\varphi}^2}2\left( 1-\frac{2\rho }{\rho _c}\right) .
\nonumber
\end{equation}
We can see that, near the bounce, the energy density is $\rho \approx \rho _c$ and the super inflation happens. In this paper we use the background solution
given in \cite{CMNS-PRD08} to calculate the power spectrum. The potential is
\begin{equation}
V=\rho _c-U(\varphi ).
\end{equation}
Now we define the parameter
\begin{equation}
\lambda =-\frac{U,_\varphi }U\sqrt{\frac{\rho _c}\rho } . \label{lambda}
\end{equation}
Considering the regime where $\rho _c/\rho \approx 1$, we can see $\lambda $
is a constant and by integrating $\lambda $ the $U$ part of the scalar
potential is given by
\begin{equation}
U=U_0e^{-\lambda \varphi }.  \label{scalar-poten}
\end{equation}

The scale factor undergoes a power law evolution
\begin{equation}
a(\tau )=\left( -\tau \right) ^p  \label{power-law}
\end{equation}
where $\tau $ is negative, and
\begin{equation}
p=-\frac 1{\bar{\epsilon}+1},~\bar{\epsilon}=-\frac{(U,_\varphi /U)^2}2\approx \frac{\lambda ^2}2.  \label{p}
\end{equation}
The time derivative and the potential expressed in the conformal time $\tau $
yields
\begin{equation}
\varphi ^{\prime }=\frac{\sqrt{2}}{\bar{\epsilon}+1}\frac 1\tau ,
\label{phi'-tau}
\end{equation}
and
\begin{equation}
V=\rho _c-\frac{3+\bar{\epsilon}}{(1+\bar{\epsilon})^2}\frac 1{(a\tau )^2}.
\label{V}
\end{equation}
We can see that in this solution, the parameter $p\rightarrow 0$ corresponds
to the standard fast-roll super inflation, which occurs when the field $%
\varphi $ is rolling down a steep potential. The equation of the variable $v$
is Eq.(\ref{Mukhanov v''})
\begin{equation}
v^{\prime \prime }-\Omega \nabla ^2v-\frac{z^{\prime \prime }}z=0,  \nonumber
\end{equation}
and the squared velocity of the perturbation is
\begin{equation}
c_s^2=\Omega.   \nonumber
\end{equation}
From the definition of $\Omega $ [Eq.(\ref{Omega})] we can see that $-1\leq \Omega \leq 1$. In the super inflation case, $\rho \simeq \rho _c$ and $%
\Omega \approx -1$. So the equation above turns out to be
\begin{equation}
v^{\prime \prime }+\nabla ^2v-\frac{z^{\prime \prime }}zv=0.  \nonumber
\end{equation}

We can Fourier decompose $v$ to give
\begin{equation}
v_k^{\prime \prime }-k^2v_k-\frac{z^{\prime \prime }}zv_k=0.  \nonumber
\end{equation}
Given Eq. (\ref{z}), Eq. (\ref{power-law}), and Eq(\ref{V}) we can get the explicit expression of $z$ as
\begin{equation}
z=\frac{\sqrt{2}\bar{\epsilon}}{\bar{\epsilon}+1}\frac 1p(-\tau )^p,
\nonumber
\end{equation}
and we have
\begin{equation}
v_k^{\prime \prime }-k^2v_k-\frac{p\left( p-1\right) }{\tau ^2}v_k=0.
\label{v_k-Eq1}
\end{equation}
In order to solve this equation, we have to substitute the variable $\tau $ into the new variable $\tau ^{\prime }:=i\tau $ and the above equation
becomes
\begin{equation}
\frac{d^2v_k}{d\tau ^{\prime }{}^2}+(k^2+m_{eff}^2)v_k=0,  \label{v_k-Eq2}
\end{equation}
where
\begin{equation}
m_{eff}^2=-\frac 1{\tau ^{\prime }{}^2}\left( \nu ^2-\frac 14\right)
\nonumber
\end{equation}
and
\begin{equation}
\nu ^2=p\left( p-1\right) +\frac 14,  \nonumber
\end{equation}
which is the same as the classical case. The general solution of Eq. (\ref{v_k-Eq2}) is
\begin{equation}
v_k=\sqrt{x}\left[ AH_\nu ^{(1)}(x)+BH_\nu ^{(2)}(x)\right]   \nonumber
\end{equation}
which is also the same as the classical case and $A$ and $B$ are constants and $H_\nu ^{(1)}$, $H_\nu ^{(2)}$ are Hankel's functions of the first and second
kind. However, in this case $x=-ik\tau $. At early times when the mode is deep inside the horizon, $k\gg aH$($|-ik\tau |\gg 1$) and the solution of Eq. (\ref{v_k-Eq2}) is
\begin{equation}
v_k=d_1e^{k\tau }+d_2e^{-k\tau }.  \label{v_k-LQC}
\end{equation}
We have used the original variable $\tau $ for convenience. There are two modes in Eq.(\ref{v_k-LQC}) and the solution is a composition of them. Now we
discuss the case that the subhorizon solution depends only on each of the modes, respectively, and then consider the general case which is a
composition of the two modes.

If the solution depends on the mode $e^{k\tau }$ only, it can be written as
\begin{equation}
v_{1k}=D_1e^{k\tau }.
\end{equation}
This equation is an exponent function which varies as $\tau $ changes its value, as a result it cannot be normalized at every time. To specify the
parameters $D_1$ , we choose a definite time $\tau _0$ to normalize it and the variable $v_k$ is written to be
\begin{equation}
v_{1k}=e^{-k\tau _0}e^{k\tau }.  \label{v_k-3}
\end{equation}

On the other hand, when $|x|\gg 1$, the Hankel function is
\begin{eqnarray}
H_\nu ^{(1)}(x &\gg &1)\sim \sqrt{\frac 2{\pi x}}e^{i\left( x-\frac \pi 2\nu
-\frac \pi 4\right) },  \nonumber \\
H_\nu ^{(2)}(x &\gg &1)\sim \sqrt{\frac 2{\pi x}}e^{-i\left( x-\frac \pi 2%
\nu -\frac \pi 4\right) }.  \nonumber
\end{eqnarray}
As we choose the ultraviolet limit to be Eq.(\ref{v_k-3}), we can fix the constant $B=0$ and $A=\sqrt{\frac \pi 2}e^{-k\tau _0+i(\frac \pi 2\nu +\frac %
\pi 4)}$. Therefore, we have
\begin{equation}
v_{1k}(\tau )=\sqrt{\frac{-i\pi k\tau }2}e^{-k\tau _0+i\left( \frac{\pi \nu }2+%
\frac \pi 4\right) }H_\nu ^{(1)}\left( -ik\tau \right).   \label{v_k-4}
\end{equation}
We can now look at the long wavelength limit. On superhorizon scales, $k\ll 1$ and for a specific finite time $\tau $ this corresponds to $|x|\ll 1$.
Since
\begin{equation}
H_\nu ^{(1)}(|x|\ll 1)\sim \sqrt{\frac 2\pi }2^{\nu -\frac 32}e^{-i\frac \pi %
2}x^{-\nu }\frac{\Gamma (\nu )}{\Gamma (\frac 32)},  \nonumber
\end{equation}
with this, in the superhorizon limit, the solution is
\begin{equation}
v_{1k}=e^{-k\tau _0}2^{\nu -\frac 32}(-k\tau )^{-\nu +\frac 12}\frac{\Gamma
(\nu )}{\Gamma (3/2)}.  \label{v_k-5}
\end{equation}

On the other hand, if we choose the ultraviolet to be the mode $e^{-k\tau }$, we have to set the constant $A=0$ and $B=\sqrt{\frac \pi 2}e^{k\tau
_0-i\left( \frac \pi 2\nu +\frac \pi 4\right) }$. In this case
\begin{equation}
v_{2k}=\sqrt{\frac{-i\pi k\tau }2}e^{k\tau _0-i\left( \frac \pi 2\nu +\frac \pi %
4\right) }H_\nu ^{(2)}\left( -ik\tau \right).   \label{v_k-6}
\end{equation}
Since
\begin{equation}
H_\nu ^{(2)}(x\ll 1)\sim -\sqrt{\frac 2\pi }2^{\nu -\frac 32}e^{i\pi \nu
}e^{-i\frac \pi 2}\left( ik\tau \right) ^{-\nu }\frac{\Gamma (\nu )}{\Gamma
(3/2)},  \nonumber
\end{equation}
we can see in the superhorizon limit
\begin{equation}
v_{2k}=2^{\nu -\frac 32}\frac{\Gamma (\nu )}{\Gamma (3/2)}e^{k\tau _0-i\pi \nu
+i\frac \pi 2}(-k\tau )^{-\nu +\frac 12}.  \label{v_k-7}
\end{equation}
Comparing Eq.(\ref{v_k-5}) and Eq.(\ref{v_k-7}) and neglecting the constant phase factor, we can see that in the latter equation the factor $e^{-k\tau _0}
$ is replaced by the new factor $e^{k\tau _0}$. In the general case, the solution is a composition of the two modes, so that
the variable $v_k$ is
\begin{equation}
v_k=Cv_{1k}+Dv_{2k}
\end{equation}
where $C$ and $D$ are complex constants. Here we also have to normalize it at the definite time $\tau_0$. As $v_{1k}(\tau_0)=1$ and $v_{2k}(\tau_0)=1$,  the constants $C$ and $D$ satisfy the normalization condition
\begin{equation}
|C+D|=1.\label{nomalization}
\end{equation}
From the superhorizon limit of $v_{1k}$ and $v_{2k}$ [Eq.(\ref{v_k-5}) and Eq.(\ref{v_k-7})] we have the superhorizon limit of the general case:
\begin{eqnarray}
v_k&=&Ce^{-k\tau _0}2^{\nu -\frac 32}\frac{\Gamma (\nu )}{\Gamma (3/2)}(-k\tau
)^{-\nu +\frac 12} \nonumber \\
&&+D2^{\nu -\frac 32}\frac{\Gamma (\nu )}{\Gamma (3/2)}%
e^{k\tau _0-i\pi \nu +i\frac \pi 2}(-k\tau )^{-\nu +\frac 12},
\end{eqnarray}
where $C$ and $D$ satisfy Eq.(\ref{nomalization}).

From Eq.(\ref{R}), ${\cal R}$ can be written as
\begin{equation}
{\cal R}=\frac{\bar{\epsilon}+1}{\sqrt{2}\bar{\epsilon}}pk^p(Ce^{-k\tau
_0}+De^{k\tau _0}),  \label{R-LQC}
\end{equation}
where we used the expansion $\nu =\frac 12-p$ as we only consider the standard super inflation. The power
spectrum is
\begin{equation}
{\cal P}_{{\cal R}}\sim k^{3+2p}(|C|^2e^{-2k\tau _0}+|D|^2e^{2k\tau
_0}+2Re(C^{*}D)).
\end{equation}
If we write $\tau _0$ to be $\frac p{a_0H _0}$, the above equation turns
out to be
\begin{equation}
{\cal P}_{{\cal R}}\sim k^{3+2p}(|C|^2e^{-2k\frac p{a_0H_0}}+|D|^2e^{2k\frac %
p{a_0H_0}}+2Re(C^{*}D)),
\end{equation}
where the subscript ``0'' indicates quantities at the conformal time ``$\tau _0$'' which we choose to normalize. We know that $k=aH$ and we can see that when
the super inflation occurs, the power spectrum varies as $k$ increases. We can consider two different limits:$p(aH/a_0H_0)\rightarrow 0$ and $p(aH/a_0H_0)\gg 1$. When $p(aH/a_0H_0)\rightarrow 0$, the factor $e^{\pm k\tau_0}\rightarrow 1$. As a result,
\begin{equation}
{\cal P}_{{\cal R}}\sim k^{\Delta n_{{\cal R}}},  \label{P_R-LQC}
\end{equation}
which is not scale invariant and has blue indices
\begin{equation}
\Delta n_{{\cal R}}=3+2p.
\end{equation}
If the other limit, $p(aH/a_0H_0)\gg 1$, is considered the power spectrum is
\begin{equation}
{\cal P}_{{\cal R}}\sim e^{-\frac p{a_0H_0}k},
\end{equation}
which is an exponentially blue spectrum.

\begin{figure}[tbp]
\includegraphics[clip,width=0.45\textwidth]{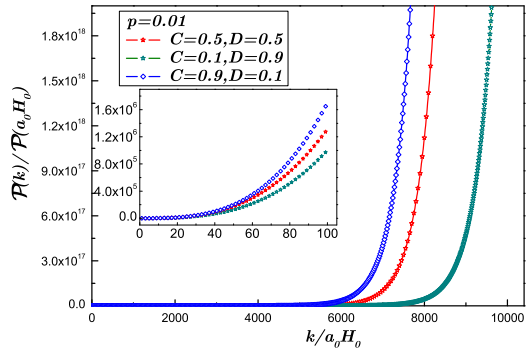}
\caption{The power spectrum with different initial conditions. The wave number $k$ varies from $a_0H_0$ to $10^4a_0H_0$. The inset depicts the region $pk/a_0H_0\leq1$. } \label{fig01}
\end{figure}

\begin{figure}[tbp]
\includegraphics[clip,width=0.45\textwidth]{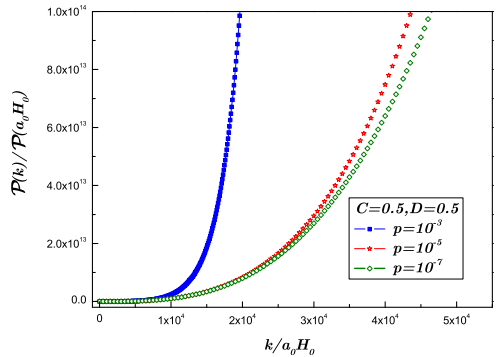}
\caption{The power spectrum with different $p$.}\label{fig02}
\end{figure}

In Fig. \ref{fig01}, different initial conditions are considered. We assign different real numbers to the constants $C$ and $D$. When the wave number increases to a great multiple, the function images of different initial conditions are the same but translated in the $x$ axis. We can also see that when $pk/a_0H_0<1$ the image approaches the cubic function. It can be verified that when $C$ and $D$ are complex numbers the graphs indicate the same signatures. When $aH$ increases sufficiently compared to the initial value $a_0H_0$, all the images turn out to be an exponential shape, which means different $k$ corresponds to different indices and the indices increase sufficiently as $k$ grows. In Fig. \ref{fig02} we can see the slope of the power spectrum curve increases as $p$ grows.

\section{\label{s4}Anisotropy}

In this section we will consider the anisotropy which is sourced by scalar perturbation. Anisotropy is described by the shear of the constant time hypersurface. In the case of the scalar perturbation metric Eq.(\ref{FRWmetric}), the shear is given by (see Appendix B of \cite{XS-PRD11})
\begin{eqnarray}
\sigma _{ij} &=&a\left[ \left( E^{\prime },_{ij}-B,_{ij}\right) -\frac 13%
\delta _{ij}\nabla ^2\left( E^{\prime }-B\right) \right]  \nonumber \\
&=&a\left( \sigma ^S,_{ij}-\frac 13\delta _{ij}\nabla ^2\sigma ^S\right),
\label{shear}
\end{eqnarray}
where the scalar shear perturbation is defined as
\begin{equation}
\sigma ^S=E^{\prime }-B.  \label{shear-perturbation}
\end{equation}
Given this we can calculate the anisotropy in the classical and quantum conditions.

\subsection{\label{s41}Anisotropy in the classical condition}

The variable ${\cal R}$ is the curvature perturbation on the comoving hypersurface, where $\delta \varphi =0$,
\begin{equation}
{\cal R}=\psi |_{\delta \varphi =0}=\psi +{\cal H}\frac{\delta \varphi }{\bar{\varphi}}.  \nonumber
\end{equation}
In this gauge the scalar mode perturbations can be represented by the gauge-invariant quantities
\begin{equation}
\left\{
\begin{array}{c}
\phi _c=\phi -\frac 1a\left( a\frac{\delta \varphi }{\bar{\varphi}^{\prime }}%
\right) ^{\prime }, \\
{\cal R}=\psi +{\cal H}\frac{\delta \varphi }{\bar{\varphi}^{\prime }}, \\
\sigma _c=\sigma ^S-\frac{\delta \varphi }{\bar{\varphi}^{\prime }}.
\end{array}
\right.   \label{gauge-invariant-qantites}
\end{equation}
In this gauge, you can verify that Eqs. (\ref{perturbation-classEq1})-(\ref{perturbation-classEq3}) turn out to be the simple form:
\begin{equation}
\nabla ^2{\cal R}+{\cal H}\nabla ^2\sigma _c=\left( {\cal H}^{\prime }-{\cal %
H}^2\right) \phi _c , \label{classEq1-1}
\end{equation}
\begin{equation}
{\cal R}^{\prime }+{\cal H}\phi _c=0,  \label{classEq1-2}
\end{equation}
\begin{equation}
{\cal R}-\phi _c+\sigma _c^{\prime }+2{\cal H}\sigma _c=0 . \label{lassEq1-3}
\end{equation}
Using Eqs.(\ref{classEq1-1}) and (\ref{classEq1-2}), and eliminating $\phi _c$, we can get
\begin{equation}
{\cal R}^{\prime }+\frac{{\cal H}}{{\cal H}^{\prime }-{\cal H}^2}\nabla ^2(%
{\cal R}+{\cal H}\sigma _c)=0.
\end{equation}
We can also eliminate $\phi _c$, $\sigma _c$ and replace ${\cal R}$ with the variable $v$ to derive the Mukhanov equation. Going to the Fourier space we
can get
\begin{equation}
\sigma _{ck}=\frac{{\cal H}^{\prime }-{\cal H}^2}{{\cal H}^2k^2}{\cal R}%
_k^{\prime }-\frac 1{{\cal H}}{\cal R}_k.
\end{equation}
Applying the solution Eq.(\ref{super-inflation}) and the expression Eq.(\ref{R_k}) to the right of the above equation, we can get
\begin{eqnarray}
\sigma _{ck} &=&\frac \lambda {2M_{pl}}\frac{\Gamma (\nu )}{\Gamma (3/2)}%
\frac 1{\sqrt{2k}}2^{\nu -\frac 32}k^{-\nu +\frac 12}(-\tau )^{-\frac 12-\nu
-p}  \nonumber \\
&&\times \left[ -\left( \frac 12-\nu -p\right) k^{-2}+(-\tau )^2\right]
\end{eqnarray}
As a comparison of the quantum case, here we consider the standard super inflation where we can expand the parameter $\nu =\frac 12-p$ and find that
the first term approaches $0$. As a result the main contribution of the anisotropy is due to the second term:
\begin{equation}
\sigma _{ck}=\frac \lambda {2\sqrt{2}M_{pl}}k^{-\frac 12+p}(-\tau ).
\end{equation}
We can also see that the anisotropy scales equally as $-\tau \sim a^{1/p}$ near the bounce, which leads to
\begin{equation}
\sigma _{cj}^i=\frac 1a\left( \sigma _{c,ij}-\frac 13\delta _{ij}\nabla
^2\sigma _c\right) \varpropto a^{\frac 1p-1}.  \label{sigma _c}
\end{equation}

We know that $p$ is negative and $|p|\ll 1$. As a result, the anisotropy sourced by the scalar perturbation in the isotropic background is much more
sensitive to the scale factor $a$, than the anisotropy derived from the standard anisotropic background which is proportional to $1/a^3$. To see how
the anisotropy grows, we estimate the size of the anisotropy:
\begin{eqnarray}
\langle (\sigma _c)^2\rangle  &=&\left\langle \frac 12\sigma ^{Sij}\sigma
_{ij}^S\right\rangle   \nonumber \\
&=&\int \frac{d^3k}{\left( 2\pi \right) ^3}\frac 1{2a^2}\left| \left(
-k_ik_j+\frac 13\delta _{ij}k^2\right) \sigma _k^S\right| ^2  \nonumber \\
&=&\int \frac 1{6\pi ^2a^2}\left| \sigma _k^S\right| ^2k^6dk
\label{sigma^2-Class}
\end{eqnarray}
and the integration is carried over the modes exited in the horizon:
\begin{equation}
\langle (\sigma ^S)^2\rangle \approx \frac{\lambda ^2(-\tau )^2}{48M_{pl}^2}%
\left. \frac{k^6}6\right| _0^{aH}\sim \frac p{M_{pl}^2}(aH)^4,
\label{Sigma^S-Class}
\end{equation}
where  we used the relationship $\tau =p/aH$.

\subsection{\label{s42}Anisotropy in LQC with holonomy corrections}

In the comoving gauge, Eq. (\ref{PerturbationEq-1}) and Eq. (\ref{PerturbationEq-2}) turn out to be
\begin{equation}
{\cal R}^{\prime }+{\cal H}\phi _c=0,  \label{PerturbationEq-1-LQC}
\end{equation}
and
\begin{equation}
\Omega \nabla ^2{\cal R}+{\cal H}\nabla ^2\sigma _c=\left( {\cal H}^{\prime
}-{\cal H}^2\right) \phi _c.  \label{PerturbationEq-2-LQC}
\end{equation}
Eliminating $\phi _c$ we can get
\begin{equation}
{\cal R}^{\prime }+\frac{{\cal H}}{{\cal H}^{\prime }-{\cal H}^2}\nabla
^2(\Omega {\cal R}+{\cal H}\sigma _c)=0.  \label{PerturbationEq-LQC}
\end{equation}
We can see that the scalar shear and the curvature perturbations source each other. We Fourier decompose the above equation and have
\begin{equation}
\sigma _{ck}=\frac{{\cal H}^{\prime }-{\cal H}^2}{k^2{\cal H}^2}{\cal R}%
_k^{\prime }-\frac \Omega {{\cal H}}{\cal R}_k.  \label{sigma _ck-1}
\end{equation}
With Eq.(\ref{power-law}) we can get
\begin{eqnarray}
\sigma _{ck} &=&\frac{(\bar{\epsilon}+1)}{\sqrt{2}\bar{\epsilon}}\frac{%
\Gamma (\nu )}{\Gamma (3/2)}2^{\nu -\frac 32}k^{-\nu +\frac 12}(-\tau
)^{-\nu -p-\frac 12}  \nonumber \\
&&\times \left( Ce^{-k\tau _0}+De^{k\tau _0-i\left( \pi \nu +\frac \pi 2%
\right) }\right)  \nonumber \\
&&\times \left[ (1+p)(\frac 12-\nu -p)k^{-2}-(-\tau )^2\right] .
\label{sigma _ck-2}
\end{eqnarray}
When we expand $\nu =\frac 12-p$, we can see that as in the classical case, the first term in the above equation approximates $0$. As a result, the main contribution is from the second term which decreases as the conformal time $\tau $ varies towards $0$:
\begin{equation}
\sigma _{ck}=-\frac{\bar{\epsilon}+1}{\sqrt{2}\bar{\epsilon}}k^p(-\tau
)(Ce^{-k\tau _0}+De^{k\tau _0}).  \label{sigma _ck-3}
\end{equation}
We can also see that the anisotropy scales equally as $-\tau \sim a^{-1/p}$ near the bounce, which is the same as the classical case, and as Eq.(\ref{sigma _c}), $\sigma _s\varpropto a^{\frac 1p-1}$.

Now we estimate the size of the anisotropy. As Eq.(\ref{sigma^2-Class}), we have
\begin{eqnarray}
&&\langle (\sigma _c)^2\rangle =\int \frac 1{6\pi ^2a^2}|\sigma _k^S|^2k^6dk=\frac{(\bar{\epsilon}+1)^2}{12\pi ^2\bar{\epsilon}^2}(-\tau )^2
\nonumber \\
&&\times \int k^6\left[|C|^2e^{-2k\tau _0}+|D|^2e^{2k\tau _0}+2Re(C^{*}D)\right]dk,
\nonumber \\
\label{sigma^2-LQC}
\end{eqnarray}
where in the last step we have neglected the order of $p$. The integration is carried over the modes that exit the horizon; one can get
\begin{eqnarray}
&&\langle (\sigma ^S)^2\rangle   \nonumber \\
&=&\frac{\left( \bar{\epsilon}+1\right) ^2}{12\pi ^2\bar{\epsilon}^2}\left\{
-\frac{\left( a_0H_0\right) ^2}{aH}\left( |C|^2e^{-2p\frac{aH}{a_0H_0}%
}+|D|^2e^{2p\frac{aH}{a_0H_0}}\right) \right.   \nonumber \\
&&\times \left[ \frac 32\left( aH\right) ^4+\frac{15}2\left( \frac{a_0H_0}p%
\right) ^2\left( aH\right) ^2+\frac{45}4\left( \frac{a_0H_0}p\right)
^4\right]   \nonumber \\
&&+p\frac{a_0H_0}{\left( aH\right) ^2}\left( |C|^2e^{-2p\frac{aH}{a_0H_0}%
}-|D|^2e^{2p\frac{aH}{a_0H_0}}\right)   \nonumber \\
&&\times \left[ -\frac 12\left( aH\right) ^6-\frac{15}2\left( \frac{a_0H_0}p%
\right) ^2\left( aH\right) ^4\right.   \nonumber \\
&&\left. -\frac{45}4\left( \frac{a_0H_0}p\right) ^4\left( aH\right) ^2\frac{%
45}8\left( \frac{a_0H_0}p\right) ^6\right]   \nonumber \\
&&\left. +\frac 27Re(C*D)p^2\left( aH\right) ^5\right\}.   \label{sigma^S-LQC}
\end{eqnarray}

\begin{figure}[tbp]
\includegraphics[clip,width=0.45\textwidth]{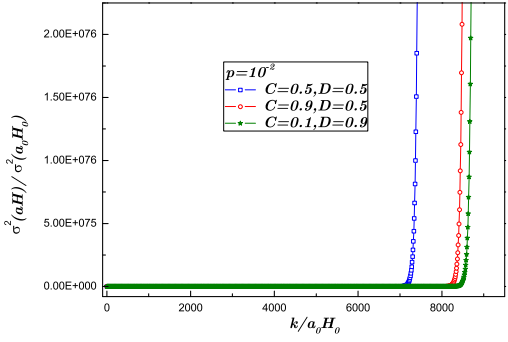}
\caption{Anisotropy growing with different initial conditions.}\label{fig1}
\end{figure}
\begin{figure}[tbp]
\includegraphics[clip,width=0.45\textwidth]{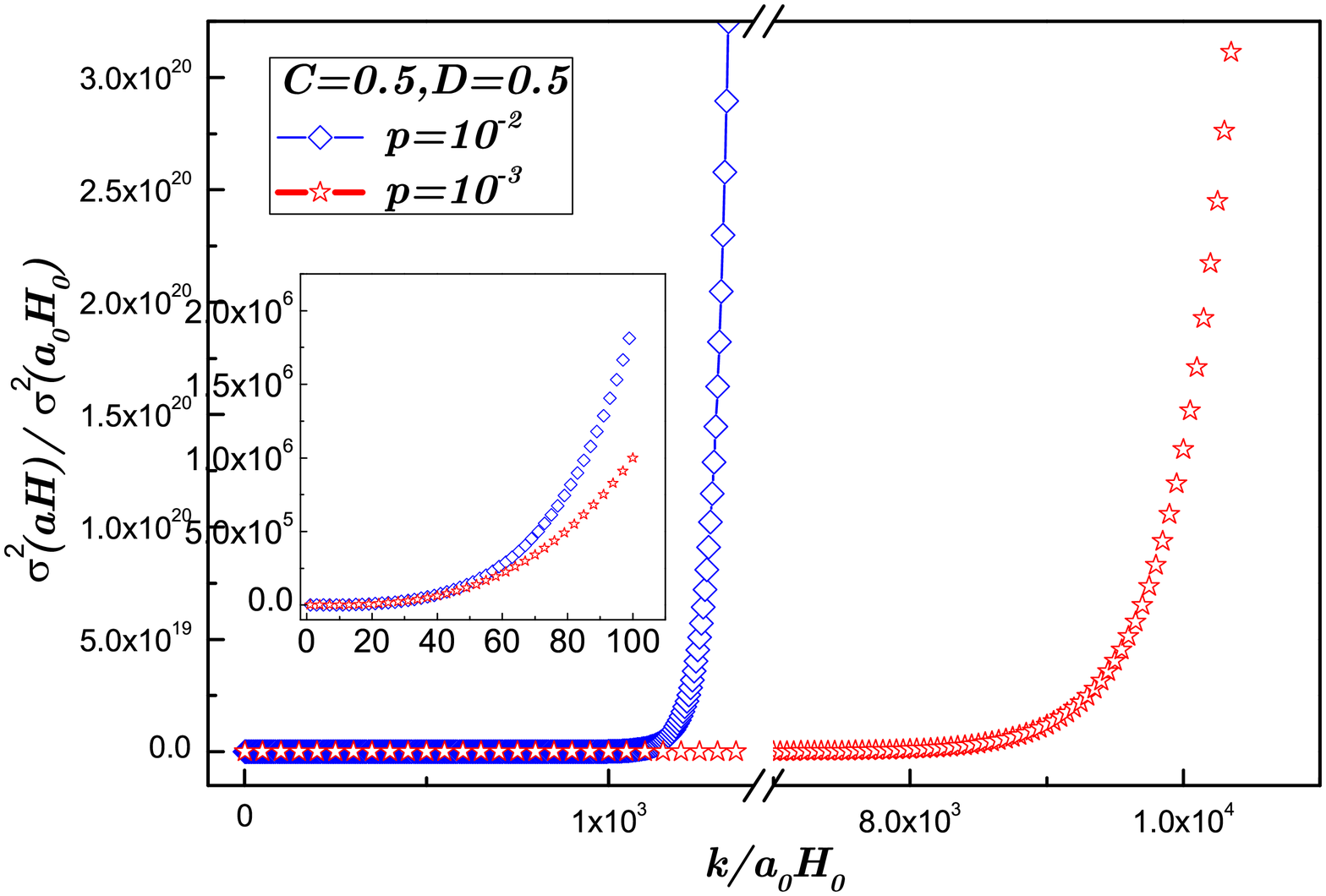}
\caption{Anisotropy growing with different values of $p$. The wave number varies from $a_0H_0$ to $10^4a_0H_0$. The inset depicts the region where the wave number varies from $a_0H_0$ to $100 a_0H_0$. }\label{fig2}
\end{figure}
The images of how $\sigma^2$ varies with $aH$ are shown in Fig. \ref{fig1} and Fig. \ref{fig2}. We can see from Fig. \ref{fig1} that, as in the power spectrum case, the magnitude of the anisotropy increases exponentially as $aH$ increases to  great large multiples compared to the initial value $a_0H_0$. Different initial conditions only translate the images in the $k$ axis but do not change the shape. We can also see from Fig. \ref{fig2} that when the value of $p$ increases, the slopes of the curves also increase.

\section{\label{s5}Discussion and conclusions}
In this paper we have studied the power spectrum and anisotropy in the super inflation epoch, respectively, in the classical condition and in LQC with holonomy corrections and the gauge-invariant scalar mode perturbation.

In the classical condition, the super inflation is driven by matters that violate the null energy condition. In this case the super inflation epoch can produce a scale-invariant power spectrum with small red tilt by choosing the parameter $p\rightarrow-1$ in the super inflation solution. However, this epoch is not the standard super inflation where the Hubble parameter $H$ grows rapidly with a rate that is much faster than the growth rate of $a$. In other words, if the power spectrum is scale invariant, the scale factor $a$ also has to grow rapidly with the rate compared to the rate of the growth of $H$. On the other hand, if the epoch is a standard super inflation, the power spectrum is not scale invariant and has blue indices.

In contrast to the classical condition, in LQC with gauge-invariant holonomy corrections, the power spectrum never has a scale-invariant form and has blue indices no matter how we choose the parameter $p$ in the super inflation solution.

In addition to the difference from the results in the classical condition, this is also different from previous investigations in LQC. The power spectrum of super inflation has already been studied in the form of LQC in previous calculations.  In \cite{MNN-PRD08}, the primordial spectrum is calculated with inverse-volume corrections and in \cite{CMNS-PRD08} the holonomy corrections are also considered.  However, these investigations are based on the perturbation theory which is not anomaly free. Besides that, in these cases the background spacetime is approximately unperturbed and the whole perturbation is due to the scalar field. With this approximation the power spectrum can be scale invariant when the parameter $p\rightarrow0$, which corresponds to the standard fast-roll super inflation. The study in this paper shows that, with the gauge-invariant form of perturbation, the scale-invariant power spectrum can never be achieved. In our investigation, we also find that the indices of the spectrum are not a constant but grow with the wave number $k$. When $k$ grows to a sufficient multiple of the beginning $a_0H_0$, the spectrum approaches an exponential function with the wave number $k$.

Then we calculated the anisotropy originated from the perturbation in classical condition and in LQC, both of which are in the standard super inflation. We find out that in both cases the anisotropy increases when $aH$ grows during inflation. In the classical condition, the anisotropy grows as $(aH)^4$. In the LQC case, the perturbation $\mathcal{R}$ grows exponentially, the anisotropy produced by the perturbation also grows exponentially, which is much faster than the classical case.

As the parameter $\Omega$ is negative during super inflation, the perturbation is unstable during this period. This causes the exponentially blue power spectrum of curvature perturbations and the exponentially increasing anisotropy. The power spectrum we calculated here is not fit to the one observed today which is nearly scale invariant. This means that the power spectrum observed today is not produced by the super inflation epoch. In LQC, when the energy density is $\frac{1}{2}\rho_c<\rho<\rho_c$, the super inflation happens, after which the standard inflation follows. The observable today is very likely to be originated to the standard inflation epoch. When the standard inflation happens in LQC, $\frac{\rho}{\rho_c}\sim\frac{1}{2}$, which corresponds to the parameter $\Omega\sim0$. In this case the inflation epoch is also different from the classical case where $\Omega=1$. The transition phase from super inflation to standard slow-roll inflation is investigated in \cite{Sadjadi}. The calculation of the power spectrum produced by this period is not of concern for this paper. In  \cite{BC-JCAP11.1},\cite{BCT-PRL11},\cite{BC-JCAP11.2}, and \cite{Calcagni} the inflationary power spectrum for
gauge-invariant perturbations has been computed for
inverse-volume corrections. However, the calculation of the power spectrum with holonomy corrections in the anomaly-free form is still an open issue and requires future investigations. The super inflation is a period just after the "big bounce", when this epoch happens the Hubble horizon contracts and the inhomogeneities of wavelength larger than the horizon is "frozen". In the following standard inflation, the structure produced by super inflation expands as $a$ increases, and as a result, the structure has the longest wavelength compared to the structure produced by the standard inflation epoch. The observation of the power spectrum may be out of reach today, but the exponentially blue indices are still a signature of the structure of that period.

The anisotropy observed today is very small, but, according to our calculations, the anisotropy during the super inflation increases exponentially as $aH$ grows. Since experiments from standard inflation suggests that approximately 60 $e$ folds of growth of $aH$ are required for consistency with observations, it seems disappointing that the resultant anisotropy during super inflation may be unacceptable. However, all the $e$ folds are not produced by the super inflation epoch. In the following standard inflation epoch the scale factor $a$ changes rapidly as $H$ remains nearly constant, which can also make $aH$ grow sufficiently. If the super inflation epoch is short enough and the growth of $aH$ is not very enormous, the divergence of anisotropy is not a problem. On the other hand, even if the anisotropy produced by the super inflation epoch is a very large quantity, the following inflation epoch is very likely to cancel it.

\acknowledgments
This work was supported by the National Natural Science Foundation of China (Grant No. 11175019 and No. 11235003).


\begin{thebibliography}{99}
\bibitem{Starobinsky}A. A. Starobinsky, Phys. Lett. B {\bf 91}, 99 (1980).
\bibitem{Guth}A. H. Guth, Phys. Rev. D {\bf 23}, 347 (1981).
\bibitem{Albrecht-Steinhardt}A. Albrecht and P. J. Steinhardt, Phys. Rev. Lett. {\bf 48}, 1220 (1982).
\bibitem{Hawking-Moss}S. W. Hawking and I. G. Moss, Phys. Lett. B {\bf 110}, 35 (1982).
\bibitem{Linde}A. D. Linde, Phys. Lett. B {\bf 108}, 389 (1982); {\bf 129}, 177 (1983).
\bibitem{Linde-Lyth}A. R. Liddle and D. H. Lyth, {\it Cosmological Inflation and Large-scale Structure} (Cambridge University Press, Cambridge, England, 2000).

\bibitem{lqg1}  T. Thiemann, {\it Introduction to Modern Canaoical Quantum General Relativity}, (Cambridge University Press, Cambridge, England,2007).

\bibitem{lqg2}  C. Rovell, {\it Quantum Gravity}, (Cambridge University Press, Cambridge, England,2004).

\bibitem{lqg3}  A. Ashtekar and J. Lewandowski, Classical Quantum Gravity {\bf 21}, R53 (2004).
\bibitem{B1-B4}  M. Bojowald, Classical Quantum Gravity {\bf 17}, 1489 (2000); {\bf 17}, 1509 (2000); {\bf 18}, 1055 (2001); {\bf 18}, 1071 (2001).
\bibitem{Rovelli-LivingRevRel}C. Rovelli, Living Rev. Relativity {\bf 1}, 1 (1998), http://www.livingreviews.org/lrr-1998-1.
\bibitem{Thiemann-Lect}T. Thiemann, Lect. Notes Phys. {\bf 631}, 41 (2003).
\bibitem{Bojowald-grqc}M. Bojowald, arXiv: gr-qc/0505057.
\bibitem{Bojowald-PRL-01}M. Bojowald, Phys. Rev. Lett. {\bf 86}, 5227 (2001).
\bibitem{Bojowald-PRD-01}M.Bojowald, Phys. Rev. D {\bf 64}, 0804018 (2001)
\bibitem{Bojowald-CQG-02}M.Bojowald, Classical Quantum Gravity {\bf 19},5113 (2002).
\bibitem{Vandersloot-PRD-05}K. Vandersloot, Phys. Tev. D {\bf 71}, 103506 (2005).
\bibitem{APS-PRL06-PRD06}A. Ashtekar, T. Pawlowski, and P. Sigh, Phys. Rev. Lett. {\bf 96},141301 (2006); Phys. Rev. D {\bf 74}, 084003 (2006).

\bibitem{APSV-PRD07}A. Ashtekar, T. Pawlowski, P. Sigh, and K. Vandersloot, Phys. Rev. D {\bf 75}, 024035 (2007).
\bibitem{Vandersloot-PRD07}K. Vandersloot, Phys. Rev. D {\bf 75}, 023523 (2007).
\bibitem{CVZ-PRD07}A. Corichi, T. Vukasinac, and J. A. Zapata, Phys. Rev. D {\bf 76}, 044016 (2007).

\bibitem{Bojowald-PRL02}M. Bojowald, Phys. Rev. Lett. {\bf 89}, 261301 (2002).
\bibitem{BV-PRD03}M. Bojowald and K. Vandersloot, Phys. Rev. D {\bf 67}, 124023 (2003).
\bibitem{DH-PRL05}G. Date and G. M. Hossain, Phys. Rev. Lett. {\bf 94}, 011301 (2005).
\bibitem{lqcs}  M. Bojowald, M. Kagan, and P. Singh, Phys. Rev. D {\bf 74}, 123512 (2006).
\bibitem{lqcv}  M. Bojowald and G. M. Hossain, Classical Quantum Gravity {\bf 24}, 4801 (2007).
\bibitem{lqct}  M. Bojowald and G. M. Hossain, Phys. Rev. D {\bf 77}, 023508 (2008).
\bibitem{lqcg1}  M. Bojowald, G. M. Hossain, and M. Kagan and S. Shankaranarayanan, Phys. Rev. D {\bf 78}, 063547 (2008).
\bibitem{BHKS-PRD09}M. Bojowald, G. M. Hossain, M. Kagan, and S. Shankaranarayanan, Phys. Rev. D {\bf 79}, 043505 (2009); {\bf 82} 109903(E) (2010).

\bibitem{Wu-Ling}J.P.Wu and Y.Ling, J. Cosmol. Astropart. Phys. {\bf 05} (2010) 026.
\bibitem{CMNS-PRD08}E. J. Copeland, D. J. Mulryne, N. J. Nunes, and M. Shaeri, Phys. Rev. D {\bf 77}, 023510 (2008).
\bibitem{afh} Thomas Cailleteau, Jakub Mielczarek, Aurelien Barrau and Julien Grain, Classical Quantum Gravity {\bf 29}, 095010 (2012).
\bibitem{KOST-PRD01}J. Khoury, B. A. Ovrut, P. J. Steinhardt, and N. Turok, Phys. Rev. D {\bf 64}, 123522 (2001).
\bibitem{XS-PRD11}BingKan Xue and Paul J. Steinhardt, Phys. Rev. D {\bf 84}, 083520 (2011).
\bibitem{MNN-PRD08}David J. Mulryne and Nelson J. Nunes, Phys. Rev. D {\bf 77}, 023510 (2008).
\bibitem{Sadjadi}H. M. Sadjadi, arXiv: 1205.1974v2
\bibitem{BC-JCAP11.1}M. Bojowald, G. Calcagni,  J. Cosmol. Astropart. Phys. {\bf 1103} (2011) 032.
\bibitem{BCT-PRL11}M. Bojowald, G. Calcagni, and S. Tsujikawa, Phys. Rev. Lett. {\bf 107}, 211302 (2011).
\bibitem{BC-JCAP11.2}M. Bojowald, G. Calcagni, and S. Tsujikawa, J. Cosmol. Astropart. Phys. {\bf 11} (2011) 046
\bibitem{Calcagni}G. Calcagni, arXiv: 1209.0473v1



\end{thebibliography}
\end{document}